\documentclass[11pt]{article}
\usepackage{epsfig}
\usepackage{graphicx}
\usepackage{amsmath}
\usepackage{amssymb}
\usepackage{amsxtra}
\usepackage{amsfonts,amsthm}
\def\Journal#1#2#3#4{{#1} {\bf #2}, #3 (#4)}

\def\IJMPA{{ Int. J. Mod. Phys.}  A}

\def\BWP{ Bled Workshops in Physics}

\def\({\left(}
\def\){\right)}

\def\beq{\begin{equation}}
\def\eeq{\end{equation}}
\def\lsim{\mathrel{\rlap{\lower4pt\hbox{\hskip1pt$\sim$}}
    \raise1pt\hbox{$<$}}}         

\def\gsim{\mathrel{\rlap{\lower4pt\hbox{\hskip1pt$\sim$}}
    \raise1pt\hbox{$>$}}}         
\title{Nonstandard cosmologies from physics beyond the Standard model}
\author{
M.Yu.~Khlopov $^{a,b}$\footnote{khlopov@apc.univ-paris7.fr}
}
\begin{document}
\maketitle
\begin{flushleft}{{\it $^{a}$ \small National Research Nuclear University "MEPHI" (Moscow Engineering Physics
Institute), 115409 Moscow, Russia \\
Centre for Cosmoparticle Physics ``Cosmion'' 115409 Moscow, Russia
\\ $^{b}$ \small APC laboratory 10, rue Alice Domon et Lonie Duquet
75205 Paris Cedex 13, France}}
\end{flushleft}
\begin{abstract}
The modern cosmology is based on inflationary models with baryosynthesis and dark matter/energy.It implies extension of particle symmetry beyond the Standard model. Studies of physical basis of the modern cosmology combine direct searches for new physics at accelerators with its indirect non-accelerator probes, in which cosmological consequences of particle models play important role. The cosmological consequences of particle models inevitably go beyond the 'standard' cosmological $\Lambda$CDM model and some possible feature of such 'nonstandard'cosmological scenarios is the subject of the present brief review.
\end{abstract}

\section{Introduction}
The now standard $\Lambda$CDM cosmological model involving inflation, baryosynthesis and dark matter/energy implies physics beyond the Standard model (BSM) of elementary particles. However, particle models, predicting new physics, can hardly reduce their cosmological consequences to these basic elements of the Standard cosmological model. One can expect that they can give rise to additional model dependent signatures of new physics and to corresponding non-Standard features of the cosmological scenario. Therefore any model that pretends to be a physical basis for the modern cosmology should be studied in more details in order to reveal such non-Standard cosmological features and we discuss in the present brief review some possible signatures for non-Standard cosmological scenarios.

The extension of the Standard model can be developed in either up-down or bottom-up direction. In the first case (as e.g. in the approach of \cite{Norma}) the overwhelming framework is proposed, which embeds the Standard model and involves new physics beyond it. To be realistic this new physics should provide inflation, baryosynthesis and dark matter. To be true it should be falsifiable, predicting model dependent signatures for its probe.

The bottom-up approach, motivated by internal problems and incompleteness of the Standard model, generally involves smaller number of parameters of new physics, than it inevitably takes place in the up-down case. It makes possible to study in more details such models and, in particular, their cosmological and astrophysical impact.

We'd like to consider physical motivations and effects of multi-component dark matter (Section \ref{multi}), as well as of primordial nonlinear structures from phase transitions in early Universe and from nonhomogeneous baryosynthesis, reflected in its extreme form in the existence of antimatter domains and antimatter objects in the baryon asymmetric Universe (Section \ref{Structures}). The importance of the account for non-standard cosmological features of physics beyond the Standard model is stressed in the conclusive Section \ref{Conclusions}.

 \section{Multi-component dark matter}\label{multi}
 Stability of elementary constituents of matter reflects the fundamental symmetry of microworld, which prevents decays of the lightest particles that possess this symmetry. So atoms are stable because electrons cannot decay owing to electric charge conservation, while protons that can in principle decay are very longliving due to the conservation of the baryon number. Physics beyond the Standard model, extending particle symmetry, involves new conservation laws that leads to stability of new forms of matter that can play the role of dark matter candidates.

 In the simplest case extension of particle symmetry can lead to only one new conservation law corresponding to a single dark matter candidate. Most popular scenarios assume that this candidate is a Weakly Interacting Massive Particle (WIMP) having strong motivation in the so called "WIMP miracle": the calculated frozen out abundance of WIMPs with mass around several tens - several hundreds GeV provides their contribution in the total density of the Universe that can explain observed dark matter density.
 The simplest WIMP scenario has an advantage to be checked in the combination of cosmological, astrophysical and physical effects. WIMP annihilation to ordinary particles not only determines their frozen out concentration but also should lead to contribution of energetic products of this annihilation to the fluxes of cosmic rays and cosmic gamma radiation. It supports indirect searches for dark matter following the original idea of \cite{ZKKC}. The same process viewed in t-channel corresponds to WIMP scattering on ordinary matter, and viewed from the opposite side to creation of WIMPs in collisions of matter particles. The former motivates direct searches for dark matter, while the latter challenges WIMP searches at accelerators and colliders. WIMP paradigm found physical motivation in supersymmetric models that were considered as the mainstream in studies of physics beyond the Standard model.

 However, direct searches for dark matter have controversial results and though their interpretation in the terms of WIMPs is still not ruled out \cite{DAMARev},
a more general approach to a possible solution of the dark matter problem is appealing. Here we'll discuss some possible nontrivial forms of cosmological dark matter that naturally follow from physics beyond the Standard Model. These examples reflect various features of multi-component dark matter either by compositeness of its species or by co-existence of various dark matter candidates.
\subsection{Composite dark matter and OHe cosmology}
 In the same way as the ordinary matter is composed by atoms, which consist of electrically charged electrons and nuclei, bound by Coulomb forces, new electrically charged stable particles may be bound by ordinary Coulomb field in the dark atoms of the dark matter. The electrically charged constituents of dark atoms may be not only elementary particles, but can be composite objects, as are ordinary nuclei and nucleons. The problem of stable electrically charged particles is that bound with electrons, such particles with charges +1 and +2 form anomalous isotopes of hydrogen and helium. In particular, the idea of stable charged particles bound in neutral dark atoms
put forward by Sheldon Glashow in his sinister model \cite{Glashow} found the unrecoverable problem of anomalous istotope overproduction, revealed in \cite{BKSR1}. It is impossible to realize the dark atom scenario in any model predicting stable +1 and -1 charged species. The former inevitably bind with electrons, forming anomalous hydrogen directly, while the latter bind first with primordial helium in  +1  charged ions, which in turn form anomalous hydrogen.

Starting from 2006 the solutions of dark atom scenario were proposed \cite{invention,KK1,ac,5g,spectro,spectro1,DMRev,DDMRev}, in which stable -2 charged species are bound with primordial helium in neutral OHe atoms, which play important catalyzing role in reduction of all the undesirable positively charged heavy species that can give rise to anomalous isotopes. Moreover OHe atoms can be a candidate for composite dark matter, dominating in the matter density of the Universe. Such candidates for dark matter should consist of
negatively doubly-charged heavy (with the mass $\sim 1$ TeV) particles, which are called O$^{--}$, coupled
to primordial helium. Lepton-like technibaryons, technileptons, AC-leptons or clusters of
three heavy anti-U-quarks of 4th generation with strongly suppressed hadronic
interactions are examples of such O$^{--}$ particles (see \cite{invention,KK1,ac,spectro,spectro1,DMRev,DDMRev} for
a review and for references). Another direction of composite dark matter scenario is to consider
neutral stable heavy quark clusters as it is proposed in the approach of \cite{Norma}. However, even in this approach heavy stable -2 charged clusters ($\bar u_5 \bar u_5 \bar u_5$) of stable antiquarks $\bar u_5$ of 5th generation can also find their physical basis \cite{5g}.

As it was qualitatively shown earlier (see \cite{sym16,Bled15} for the latest review), the transfer function of density perturbations of OHe dark matter has specific Warmer-than-Cold features, reflecting its composite nature and the nuclear cross sections for OHe elastic collisions with nuclei.

The cosmological and astrophysical effects of such composite dark matter (dark atoms of
OHe) are dominantly related to the helium shell of OHe and involve only one parameter
of new physics $-$ the mass of O$^{--}$.

If dark matter can bind to
normal matter, the observations could come from radiative capture of thermalized OHe and could depend on the detector composition and temperature.
In the matter of the underground detector local concentration of OHe is determined by the equilibrium between the infalling cosmic OHe flux and its
diffusion towards the center of Earth. Since the infalling flux experiences annual changes due to Earth's rotation around Sun, this local OHe
concentration possess annual modulations.

 The positive results of the DAMA/NaI and
DAMA/LIBRA
experiments are then explained by the annual modulations of the rate of radiative capture of OHe
by sodium nuclei. Such radiative capture to a low energy OHe-nucleus bound state is possible only for intermediate-mass nuclei:
this explains the negative results of the XENON100 and LUX experiments. The rate of this capture can be calculated by the analogy with
radiative capture of neutron by proton, taking into account the scalar and isoscalar nature of He nucleus, what makes possible only E1 transition
with isospin violation in this process. In the result this rate is
proportional to the temperature (to the square of relative velocity in the absence of local thermal equilibrium): this leads to a suppression of this effect in cryogenic detectors, such as CDMS.

The timescale of OHe collisions in the Galaxy exceeds the age of the Universe, what proves that
the OHe gas is collisionless. However the rate of such collisions is nonzero and grows in the regions
of higher OHe density, particularly in the central part of the Galaxy, where these
collisions lead to OHe
excitations. De-excitations of OHe with pair production in E0 transitions can explain the
excess of the positron-annihilation line, observed by INTEGRAL in the galactic bulge \cite{DMRev,DDMRev,CDM,KK2,KMS,CKW,CKW3}.
The calculated rate of collisions and OHe excitation in them strongly depends on OHe density and relative velocity and
the explanation of positron excess was found to be very sensitive to the dark matter density in the central part of Galaxy, where baryonic matter dominates
and theoretical estimations are very uncertain. The latest analysis of dark matter distribution favors more modest values of
dark matter central density, what fixes the explanation of the
excess of the positron-annihilation line by OHe collisions and de-excitation in a very narrow range of the mass of O$^{--}$ near 1.25 TeV.

In a two-component dark atom model, based on Walking Technicolor, a sparse WIMP-like component of atom-like
state, made of positive and negative doubly charged techniparticles, is present together with the dominant OHe dark atom and the decays of
doubly positive charged techniparticles to pairs of same-sign leptons can explain the excess of
high-energy cosmic-ray positrons, found in PAMELA and AMS02 experiments \cite{laletin}.
This explanation is possible for the mass of decaying +2 charged particle below 1 TeV and depends on the branching ratios of leptonic channels.
Since even pure lepton decay channels are inevitably accompanied by gamma radiation the important constraint on this model follows from the measurement of cosmic gamma ray background in FERMI/LAT experiment.
It may be shown that the constraints on this background may be satisfied if the decaying component is distributed in disc and not in halo, what implies more sophisticated self-interacting nature of this component. In fact, a serious problem for any source of cosmic poitrons distributed in halo and not concentrated in the disc.

The crucial problem of OHe scenario is the existence of a dipole barrier in OHe nuclear interaction.
The scenario in which such a barrier does not appear was considered in  \cite{CKW4} and The  over-abundance of anomalous isotopes in terrestrial matter seems to be unavoidable in this case..

This makes the full solution of OHe nuclear physics, started in \cite{CKW2}, vital.
The answer to the possibility of the creation of a dipole Coulomb barrier in OHe interaction with nuclei is crucial. Indeed, the model cannot work if no repulsive interaction appears at some distance between

The problem of the Earth's shadowing represents another potential problem for OHe scenario.
The terrestrial matter is opaque for OHe, what should inevitably lead to an effect of Earth matter shadowing for
the OHe flux and corresponding  diurnal modulation, constrained in DAMA/LIBRA experiment  \cite{DAMAshadow}.
The OHe model involves only one parameter of new physics - mass of O$^{--}$ and its cosmological effects are related to nuclear and atomic physics, being within the Standard model, but even in this case many principal features of cosmological consequences remain unclear.
\subsection{Mirror atoms}\label{mirror}
Mirror particles, first proposed by T. D. Lee and C. N. Yang in Ref. \cite{LeeYang} to restore equivalence of left- and right-handed co-ordinate systems in the presence of P- and C- violation in weak interactions, should be strictly symmetric by their properties to their ordinary twins. After discovery of CP-violation it was shown by I. Yu. Kobzarev, L. B. Okun and I. Ya. Pomeranchuk in Ref. \cite{KOP} that mirror partners cannot be associated with antiparticles and should represent a new set of symmetric partners for ordinary quarks and leptons with their own strong, electromagnetic and weak mirror interactions. It means that there should exist mirror quarks, bound in mirror nucleons by mirror QCD forces and mirror atoms, in which mirror nuclei are bound with mirror electrons by mirror electromagnetic interaction \cite{ZKrev,FootVolkas}. If gravity is the only common interaction for ordinary and mirror particles, mirror matter can be present in the Universe in the form of elusive mirror objects, having symmetric properties with ordinary astronomical objects (gas, plasma, stars, planets...), but causing only gravitational effects on the ordinary matter \cite{Blin1,Blin2}.

Even in the absence of any other common interaction except for gravity, the observational data on primordial helium abundance and upper limits on the local dark matter seem to exclude mirror matter, evolving in the Universe in a fully symmetric way in parallel with the ordinary baryonic matter\cite{Carlson,FootVolkasBBN}. The symmetry in cosmological evolution of mirror matter can be broken either by initial conditions\cite{zurabCV,zurab}, or by breaking mirror symmetry in the sets of particles and their interactions as it takes place in the shadow world\cite{shadow,shadow2}, arising in the heterotic string model. We refer to Refs.
\cite{newBook,OkunRev,Paolo} for current review of mirror matter and its cosmology.

Mirror matter in its fully symmetric implementation doesn't involve new parameters of new physics, since all the parameters of mirror particles and their interactions are by construction strictly equal to the corresponding values of their ordinary partners. However, though there is no common interactions between ordinary and mirror matter except for gravity, just the presence of mirror particles in the same space-time with ordinary matter causes contradictions with observations in a strictly symmetric mirror matter cosmology.
\subsection{Unstable particles}
 \label{unstable}
 The next to lightest particle that possess a new conserved charge may be sufficiently longliving to retain some observable trace in the Universe.

Primordial unstable particles with the lifetime, less than the age
of the Universe, $\tau < t_{U}$, can not survive to the present
time. But, if their lifetime is sufficiently large to satisfy the
condition $\tau \gg (m_{Pl}/m) \cdot (1/m)$, their existence in
early Universe can lead to direct or indirect traces\cite{khlopov7}.

Weakly interacting particles, decaying to invisible modes, can influence Large Scale Structure formation.
Such decays prevent formation of the structure, if they take place before the structure is formed.
Invisible products of decays after the structure is formed should contribute in the cosmological dark energy.
The Unstable Dark matter scenarios\cite{UDM,UDM1,UDM2,UDM3,berezhiani4,berezhiani5,TSK,GSV,Sakharov1} implied weakly interacting particles that form the structure on the matter dominated stage and then decay to invisible modes after the structure is formed.

Cosmological
flux of decay products contributing into the cosmic and gamma ray
backgrounds represents the direct trace of unstable particles\cite{khlopov7,sedelnikov}. If
the decay products do not survive to the present time their
interaction with matter and radiation can cause indirect trace in
the light element abundance\cite{khlopovlinde3,khlopov3,khlopov31,DES} or in the fluctuations of thermal
radiation\cite{UDM4}.

If the particle lifetime is much less than $1$s the
multi-step indirect traces are possible, provided that particles
dominate in the Universe before their decay. On the dust-like
stage of their dominance black hole formation takes place, and the
spectrum of such primordial black holes traces the particle
properties (mass, frozen concentration, lifetime) \cite{polnarev,khlopov0,polnarev0}.
The particle decay in the end of dust like stage influences the
baryon asymmetry of the Universe. In any way cosmophenomenoLOGICAL
chains link the predicted properties of even unstable new
particles to the effects accessible in astronomical observations.
Such effects may be important in the analysis of the observational
data.

\section{Primordial cosmological structures}\label{Structures}
\subsection{Relics of phase transitions in very early Universe}
Parameters of new stable and metastable particles are also
determined by a pattern of particle symmetry breaking. This pattern
is reflected in a succession of phase transitions in the early
Universe. First order phase transitions proceed through bubble
nucleation, which can result in black hole formation (see e.g.
Refs. \cite{kkrs} and \cite{book2} for review and references). Phase
transitions of the second order can lead to formation of topological
defects, such as walls, string or monopoles. The observational data
put severe constraints on magnetic monopole \cite{kz} and cosmic
wall production \cite{okun}, as well as on the parameters of cosmic
strings \cite{zv1,zv2}. Structure of cosmological defects can be
changed in succession of phase transitions. More complicated forms
like walls-surrounded-by-strings can appear. Such structures can be
unstable, but their existence can leave a trace in nonhomogeneous
distribution of dark matter and give rise to large scale structures
of nonhomogeneous dark matter like {\it archioles}
\cite{Sakharov2,kss,kss2}. This effect should be taken into account in the analysis of
cosmological effects of weakly interacting slim particles (WISPs) (see Ref. \cite{jaeckel} for current review) that can play the role of cold dark matter in spite of their small mass.

A wide class of particle models possesses a symmetry breaking
pattern, which can be effectively described by
pseudo-Nambu--Goldstone (PNG) field and which corresponds to
formation of unstable topological defect structure in the early
Universe (see Ref. \cite{book2} for review and references). The
Nambu--Goldstone nature in such an effective description reflects
the spontaneous breaking of global U(1) symmetry, resulting in
continuous degeneracy of vacua. The explicit symmetry breaking at
smaller energy scale changes this continuous degeneracy by discrete
vacuum degeneracy. The character of formed structures is  different
for phase transitions, taking place on post-inflationary and
inflationary stages.
\subsection{Large scale correlations of axion field}\label{axion}
At high temperatures such a symmetry breaking pattern implies the
succession of second order phase transitions. In the first
transition, continuous degeneracy of vacua leads, at scales
exceeding the correlation length, to the formation of topological
defects in the form of a string network; in the second phase
transition, continuous transitions in space between degenerated
vacua form surfaces: domain walls surrounded by strings. This last
structure is unstable, but, as was shown in the example of the
invisible axion \cite{Sakharov2,kss,kss2}, it is reflected in the
large scale inhomogeneity of distribution of energy density of
coherent PNG (axion) field oscillations. This energy density is
proportional to the initial value of phase, which acquires dynamical
meaning of amplitude of axion field, when axion mass is switched on
in the result of the second phase transition.

The value of phase changes by $2 \pi$ around string. This strong
nonhomogeneity of phase leads to corresponding nonhomogeneity of
energy density of coherent PNG (axion) field oscillations. Usual
argument (see e.g. Ref. \cite{kim} and references therein) is essential
only on scales, corresponduing to mean distance between strings.
This distance is small, being of the order of the scale of
cosmological horizon in the period, when PNG field oscillations
start. However, since the nonhomogeneity of phase follows the
pattern of axion string network this argument misses large scale
correlations in the distribution of oscillations' energy density.

Indeed, numerical analysis of string network (see review in the
Ref. \cite{vs}) indicates that large string loops are strongly suppressed
and the fraction of about 80\% of string length, corresponding to
long loops, remains virtually the same in all large scales. This
property is the other side of the well known scale invariant
character of string network. Therefore the correlations of energy
density should persist on large scales, as it was revealed in Refs.
\cite{Sakharov2,kss,kss2}.

The large
scale correlations in topological defects and their imprints in
primordial inhomogeneities is the indirect effect of inflation, if
phase transitions take place after reheating of the Universe.
Inflation provides in this case the equal conditions of phase
transition, taking place in causally disconnected regions.
\subsection{Primordial seeds for Active Galactic Nuclei}\label{AGN}
If the phase transitions take place on inflational stage new forms
of primordial large scale correlations appear. The example of
global U(1) symmetry, broken spontaneously in the period of
inflation and successively broken explicitly after reheating, was
considered in Ref. \cite{RKS}. In this model, spontaneous U(1)
symmetry breaking at inflational stage is induced by the vacuum
expectation value $\langle \psi \rangle = f$ of a complex scalar
field $\Psi = \psi \exp{(i \theta)}$, having also explicit
symmetry breaking term in its potential $V_{eb} = \Lambda^{4} (1 -
\cos{\theta})$. The latter is negligible in the period of
inflation, if $f \gg \Lambda$, so that there appears a valley
relative to values of phase in the field potential in this period.
Fluctuations of the phase $\theta$ along this valley, being of the
order of $\Delta \theta \sim H/(2\pi f)$ (here $H$ is the Hubble
parameter at inflational stage) change in the course of inflation
its initial value within the regions of smaller size. Owing to
such fluctuations, for the fixed value of $\theta_{60}$ in the
period of inflation with {\it e-folding} $N=60$ corresponding to
the part of the Universe within the modern cosmological horizon,
strong deviations from this value appear at smaller scales,
corresponding to later periods of inflation with $N < 60$. If
$\theta_{60} < \pi$, the fluctuations can move the value of
$\theta_{N}$ to $\theta_{N} > \pi$ in some regions of the
Universe. After reheating, when the Universe cools down to
temperature $T = \Lambda$ the phase transition to the true vacuum
states, corresponding to the minima of $V_{eb}$ takes place. For
$\theta_{N} < \pi$ the minimum of $V_{eb}$ is reached at
$\theta_{vac} = 0$, whereas in the regions with $\theta_{N} > \pi$
the true vacuum state corresponds to $\theta_{vac} = 2\pi$. For
$\theta_{60} < \pi$ in the bulk of the volume within the modern
cosmological horizon $\theta_{vac} = 0$. However, within this
volume there appear regions with $\theta_{vac} = 2\pi$. These
regions are surrounded by massive domain walls, formed at the
border between the two vacua. Since regions with $\theta_{vac} =
2\pi$ are confined, the domain walls are closed. After their size
equals the horizon, closed walls can collapse into black holes.
The minimal mass of such black hole is determined by the condition
that it's Schwarzschild radius, $r_{g} = 2 G M/c^{2}$ exceeds the
width of the wall, $l \sim f/\Lambda^{2}$, and it is given by
$M_{min} \sim f (m_{Pl}/\Lambda)^{2}$. The maximal mass is
determined by the mass of the wall, corresponding to the earliest
region $\theta_{N} > \pi$, appeared at inflational stage.

  This mechanism can lead to formation
of primordial black holes of a whatever large mass (up to the mass
of AGNs \cite{AGN,DER1}, see for latest review Ref. \cite{PBHrev}). Such black
holes appear in the form of primordial black hole clusters,
exhibiting fractal distribution in space
\cite{KRS,Khlopov:2004sc,book2}. It can shed new light on the
problem of galaxy formation \cite{book2,DER1}.

The described mechanism of massive PBH clouds formation may be of special interest for the interpretation of the recently discovered gravitational wave signals from coalescence of massive black hole (BH) binaries \cite{gw1,gw2}. It naturally leads to formation of massive BH binaries within such a cloud, while the mass range of PBHs, determined by $f$ and $\Lambda$ can naturally cover the values of tens of Solar mass.
\begin{figure}
\begin{center}
\includegraphics[scale=0.7]{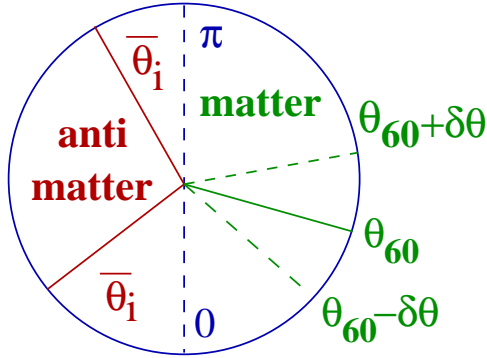}
\caption{ The inflational evolution of the
phase (taken from the Ref. \cite{ere}). The phase  $\theta_{60}$ sits in the range $[\pi ,0 ]$ at the
beginning of inflation and makes Brownian step
$\delta\theta_{eff}=H_{infl}/(2\pi f_{eff})$  at each e--fold. The
typical wavelength of the fluctuation $\delta\theta$ is equal to
$H^{-1}_{infl}$. The whole domain  $H^{-1}_{infl}$, containing phase
$\theta_{N}$ gets  divided, after one e--fold, into  $e^3$ causally
disconnected domains of radius $H^{-1}_{infl}$. Each new domain
contains almost homogeneous phase value
$\theta_{N-1}=\theta_{N}\pm\delta\theta_{eff}$. Every successive
e-fold this process repeats in every domain.}
\label{pnginfl}
\end{center}
\end{figure}
\subsection{Antimatter in Baryon asymmetric Universe?}\label{antimatter}
Primordial strong inhomogeneities can also appear in the baryon
charge distribution. The appearance of antibaryon domains in the
baryon asymmetrical Universe, reflecting the inhomogeneity of
baryosynthesis, is the profound signature of such strong
inhomogeneity \cite{CKSZ}. On the example of the model of
spontaneous baryosynthesis (see Ref. \cite{Dolgov} for review) the
possibility for existence of antimatter domains, surviving to the
present time in inflationary Universe with inhomogeneous
baryosynthesis was revealed in \cite{KRS2}.

The mechanism of
spontaneous baryogenesis \cite{Dolgov} implies the existence of a
complex scalar field $\chi =(f/\sqrt{2})\exp{(\theta )}$ carrying
the baryonic charge. The $U(1)$ symmetry, which corresponds to the
baryon charge, is broken spontaneously and explicitly. The explicit
breakdown of $U(1)$ symmetry is caused by the phase-dependent term
 \beq\label{expl} V(\theta )=\Lambda^4(1-\cos\theta ).
 \eeq
The possible baryon and lepton number violating interaction
of the field $\chi$ with matter fields can have the following
structure \cite{Dolgov} \beq\label{leptnumb} {\cal
L}=g\chi\bar QL+{\rm h.c.}, \eeq where fields $Q$ and $L$ represent
a heavy quark and lepton, coupled to the ordinary matter fields.

In the early Universe, at a time when the friction term, induced by
the Hubble constant, becomes comparable with the angular mass
$m_{\theta}=\frac{\Lambda^2}{f}$, the phase $\theta$ starts to
oscillate around the minima of the PNG potential and decays into
matter fields according to (\ref{leptnumb}). The coupling
(\ref{leptnumb}) gives rise to the following \cite{Dolgov}: as
the phase starts to roll down in the clockwise direction (Fig.~\ref{pnginfl}),  it preferentially creates
excess of baryons over antibaryons, while the opposite is true as it
starts to roll down in the opposite direction.

The fate of such antimatter regions depends on their size. If the
physical size of some of them is larger than  the critical surviving size
$L_c=8h^2$ kpc~\cite{KRS2}, they survive annihilation with surrounding matter.
Evolution of
sufficiently dense antimatter domains can lead to formation of
antimatter globular clusters \cite{GC}. The existence of such
cluster in the halo of our Galaxy should lead to the pollution of
the galactic halo by antiprotons. Their annihilation can reproduce
\cite{Golubkov} the observed galactic gamma background in the
range tens-hundreds MeV. The prediction of antihelium component of
cosmic rays \cite{ANTIHE}, accessible to future searches for
cosmic ray antinuclei in PAMELA and AMS II experiments, as well as
of antimatter meteorites \cite{ANTIME} provides the direct
experimental test for this hypothesis.

So the primordial strong inhomogeneities in the distribution of
total, dark matter and baryon density in the Universe is the new
important phenomenon of cosmological models, based on particle
models with hierarchy of symmetry breaking.

\section{Conclusions}\label{Conclusions}

As soon as physics beyond the Standard model involves new symmetries and mechanisms of their breaking, new model dependent non-standard features of the cosmological scenario should inevitably appear.
Even rather restricted list of possible examples of such features, presented here, gives the flavor of new cosmology that can come from the new physics.

The wider is the symmetry group embedding the symmetry of the Standard model of elementary particles, the larger is the list of cosmologically viable predictions that provide various probes for the considered particle model. Entering the corresponding multi-dimensional space of parameters, we simultaneously increase the set of their probes. It makes the set of equations for these parameters over-determined and provides a complete test for however extensive theoretical model.

One can conclude that the account for non-Standard cosmological scenarios in the analysis of the data of precision cosmology extends the space of cosmological parameters and provides nontrivial test for physics beyond the Standard model.

\section*{Acknowledgements}
I express my gratitude to Jean-René Cudell for kind hospitality in the University of Liege, where this contribution was completed.
The work was performed within the framework of the Center
FRPP supported by MEPhI Academic Excellence Project (contract 02.03.21.0005, 27.08.2013). The part on initial cosmological conditions was supported by the Ministry of Education and Science of Russian Federation,
project 3.472.2014/K  and on the forms of dark matter by grant RFBR 14-22-03048.

\end{document}